\documentclass[12pt]{article}
\usepackage{graphicx,epsfig,epsf,amssymb}
\usepackage{times}
\def \<{\langle}
\def \>{\rangle}

\title {\bf Evolution of gravity-dark energy\\ coupled expanding universe}
\author
{Ti-Pei Li$^{1,2}$ and Mei Wu$^{2}$}
\date{}

\begin{document}
\baselineskip 24pt
\maketitle

{\footnotesize
\begin{enumerate}
\item{Department of Physics \& Center for Astrophysics, Tsinghua University, Beijing, China}
\item{Key Laboratory of Particle Astrophysics, Institute of High Energy Physics, Chinese Academy of Sciences, Beijing, China}
\end{enumerate}
}

\noindent {\bf The acceleration of the cosmic expansion revealed by supernova observations indicates that the present universe is dominated by repulsive dark energy. It is natural that the coupling between the dark energy and the attractive gravity has to be taken into account in studying the evolution of our universe. Here we induce a dynamic equation of universe evolution based on the cosmological principal,
gravity-dark energy coupling, and energy conservation. The solutions of this dynamic equation provide a specific picture of cosmic evolution, which is well compatible with current observed evolution histories of both Hubble parameter $H(z)$ and equation of state parameter $w(z)$ of dark energy. In this picture, the already observed universe has undergone through three epochs -- the epoch of equilibrium between gravity and dark energy at cosmological redshifts $z>1$, the phase transition epoch at $z$ between about $0.5$ and $1$, and the present acceleration epoch starting from $z\simeq 0.5$.  The expected matter-dominated deceleration epoch has not been seen yet, which requires future deep observations to look back upon more early universe.
}\\

The conventional dynamic equation to describe the evolution of the universe is the Friedmann equation deduced from Einstein's field equation for general relativity under the cosmological principle, i.e., the universe is spatially homogeneous and isotropic. Milne~$^{\cite{mil34,mil35}}$ and McCrea \& Milne~$^{\cite{mcc34}}$ pointed out that, under the cosmological principle, the Friedmann equation can also be derived with Newtonian gravity -- for a homogeneous and isotropic universe, the Friedmann equation can be derived from the Poisson's equation for gravitational potential, Newton's law of motion, and conservation of energy. We now derive the dynamic equation with Milne and McCrea's approach for a gravity-dark energy coupled universe. 
 
Here we use the natural units with $c=1$. The Poisson's equation for the gravitational potential $\phi_{_m}$ of matter field with density $\rho_{_m}$ is
\[ 
\nabla^2\phi_{_m}=4\pi G\rho_{_m}\,.
\]  
The field equation of dark energy is assumed as
\[ 
\nabla^2\phi_{_\lambda}=-\lambda\,.
\]  
Choosing a common zero point for the two potentials at the origin of comoving coordinates, the potentials of gravity and dark energy at distance $R$ can be written by 
\[ 
\phi_{_m}=-\frac{4\pi G}{3}\rho_{_m}R^2
\] 
and
\[ 
\phi_{_\lambda}=\frac{\lambda}{6}R^2=\frac{4\pi G}{3}\rho_{_\lambda}R^2 
\]  
with density of dark energy $\rho_{_\lambda}=\lambda/8\pi G$. Then we get the combining potential for the two coupled fields
\[  
\phi=\phi_{_m}+\phi_{_\lambda}=-\frac{4\pi G}{3}\rho_{_d}R^2
\]   
where the differential energy density
\[  
\rho_{_d}=\rho_{_m}-\rho_{_\lambda}\,.
\]    

For an expanding universe, at age $t$ the scale factor $a(t)$ is defined by $a(t)=R(t)/R_0$, where $R(t)$ is the universe radius at $t$ and $R_0$ the present radius, and the universe volume $V=4\pi R_0^3a^3/3$. Under the rest energy conservation, the total rest energy of universe $E_{rest}=\rho V=(\rho_{_m}+\rho_{_\lambda}) V=\mbox{constant}$, then the energy density 
\[ \rho=\rho_{_m}+\rho_{_\lambda}=\rho_{_0} a^{-3} \]
with $\rho_{_0}$ being the present energy density.

An expanding universe has kinetic energy $E^{(k)}=E_{rest}\dot{R}^2/2=2\pi\rho_0R_0^5 \dot{a}^2/3$, and potential energy $E^{(p)}=-4\pi G \rho_{_d}VR^2/3=-16\pi^2GR_0^5\rho_0\rho_{_d}a^2/9$. From the conservation of mechanical energy, $E_{mech}=E^{(k)}+E^{(p)}=\mbox{constant}$,
we obtain the first evolution equation (energy conservation equation)
\begin{equation}\label{eq:f1}
\dot{a}^2=\epsilon+\frac{8\pi G}{3}\rho_{_d} a^2
\end{equation}
with the constant 
\[ \epsilon=\frac{3E_{mech}}{2\pi\rho_0R_0^5}\,. \]

At the surface of the sphere of an expanding universe with radius $R$, the force experienced by unit mass $P=P_m+P_\lambda$, where the gravitational  force $P_m=-G\rho_{_m}VR^{-2}=-4\pi G \rho_{_m} R_0 a/3$, and repulsive force from dark energy $P_\lambda=-4\pi G\rho_{_\lambda}R_0 a/3$. With  the Newton's second law of motion, $P=E_{rest}\ddot{R}$, we obtain the second evolution equation (acceleration equation)
\begin{equation}\label{eq:f2}
\ddot{a}=-\frac{4\pi G}{R_0\rho_0}\rho_{_d} a^3\,.
\end{equation}

It should be noted that the energy conservation equation\,(\ref{eq:f1}) is almost the same in form with the Friedmann equation or Friedmann-Lemaitre equation, but except that instead of the differential energy density $\rho_{_d}=\rho_{_m}-\rho_{_\lambda}$ using in Eq.\,(\ref{eq:f1}), the Friedmann equation uses only the gravitational energy density $\rho=\rho_{_m}$ and the Friedmann-Lemaitre equation uses the total energy density $\rho=\rho_{_m}+\rho_{_\lambda}.$  From the acceleration equation\,(\ref{eq:f2}) one can see that, in the gravity-dark energy coupling model, the expansion of universe is driven by the difference of gravitational energy and repulsive dark energy: In a matter dominated phase, $\rho_{_m}>\rho_{_\lambda}$ ($\rho_{_d}>0$), the expansion of universe is decelerated ($\ddot{a}<0$); in an equilibrium state, $\rho_{_m}=\rho_{_\lambda}$ ($\rho_{_d}=0$), the universe is uniformly expanding with a constant rate ($\ddot{a}=0$); and in a dark energy dominated phase, $\rho_{_m}<\rho_{_\lambda}$ ($\rho_{_d}<0$), the expansion is acceleration ($\ddot{a}>0$). 

Eliminating $\rho_{_d}$ from Eqs.\,(\ref{eq:f1}) and (\ref{eq:f2}), we finally obtain the evolution equation linking universe scale factor $a$, expansion rate $\dot{a}$ and acceleration $\ddot{a}$ 
\begin{equation}\label{eq:f3}
\ddot{a}+\mu(\dot{a}^2-\epsilon) a=0 
\end{equation}
with the constant
\[ \mu=\frac{3}{2R_0\rho_0}\,. \]

The evolution equation\,(\ref{eq:f3}), being a type of Bernoulli's equations for fluid dynamics, has three expansion solutions:
\begin{eqnarray}
\dot{a} & =& \sqrt{c_1\exp(-\mu a^2)+\epsilon} \hspace{5mm} \mbox{for}~\dot{a}>\sqrt{\epsilon} \hspace{5mm} (\ddot{a}< 0,~ \mbox{deceleration}) \hspace{16mm} (s1) \nonumber \\
\dot{a} & =& \sqrt{\epsilon} \hspace{34mm} \mbox{for}~\dot{a}=\sqrt{\epsilon} \hspace{5mm} (\ddot{a}=0, ~   \mbox{constant expansion}) \hspace{5mm} (s2) \nonumber \\
\dot{a} & =& \sqrt{\epsilon-c_2\exp(-\mu a^2)} \hspace{5mm} \mbox{for}~\dot{a}<\sqrt{\epsilon} \hspace{5mm} (\ddot{a}> 0,~ \mbox{acceleration}) \hspace{17mm} (s3) \nonumber 
\end{eqnarray}
where $c_1$ and $c_2$ are positive integral constants. The solutions (s1)-(s3) include three parameters of universe: the present radius $R_0$, the present rest energy density $\rho_{_0}$ (or total rest energy $E_{rest}=4\pi R_0^3 \rho_{_0}/3$), and the present mechanical energy density $\rho_{_{mech,0}}$ (or total mechanical energy $E_{mech}=4\pi R_0^3 \rho_{_{mech,0}}/3$). The three parameters, $R_0$, $\rho_{_0}$, and $\rho_{_{mech,0}}$ determine all possible evolutionary tracks of the universe. The integral constant $c_1$ (or $c_2$) selects a particular deceleration (or acceleration) expanding track from all possible ones.

From the solutions (s1)-(s3), we see that there in a universe exists a critical rate $\dot{a}_c=\sqrt{\epsilon}$  which determines the state of expansion. After the accelerating inflation and matter creation, the universe should enter into a matter dominated deceleration phase with a differential density $\rho_{_d}>0$  describing by the solution\,(s1) with a specific  value of $c_1$. With Eqs.\,(\ref{eq:f1}),\,(\ref{eq:f2}) and $\rho_{_\lambda}/\rho<1$, it is easy to prove that, during a deceleration phase with $\rho_{_d}>0$, the changing rate of fractional dark energy $(\frac{\rho_{_\lambda}}{\rho})'>0$, i. e., during deceleration expansion, kinetic energy is decreasing and potential energy increasing,  gravitational energy is being converting into dark energy, consequently, potential energy converting into kinetic energy. While the dark energy reaching up to be equal to the gravitational energy, $\rho_{_d}=0$, expansion rate  decreased down to the critical value, $\dot{a}=\dot{a}_c$, the universe entered the uniform expansion epoch (s2).  Once entering (s2), the constant expansion will last forever because from Eq.\,(\ref{eq:f2}) the acceleration $\ddot{a}=0$, and the universe cannot enter acceleration phase except a phase transition for transforming more gravitational energy to dark energy. After the phase transition the universe possessing  a negative differential energy density $\rho_{_d}<0$ can enter into the acceleration phase described by the solution (s3) with a specific constant $c_2$.

\begin{figure}[t]
\begin{center}
\psfig{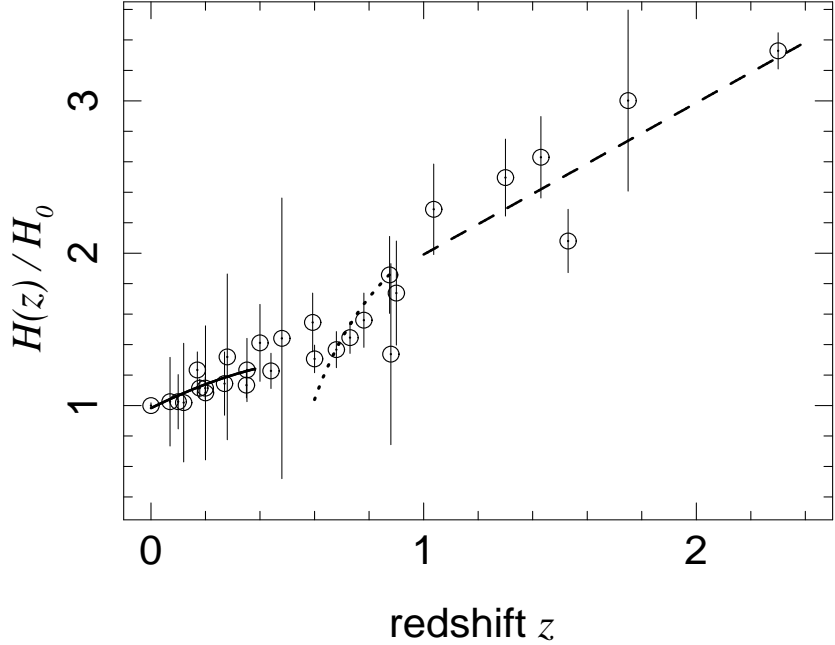}
\caption{Normalized Hubble parameter versus redshift.  Data source: $Planck$ 2013 results$^{\cite{ade13}}$ for $H_0$ and [5-11] compiled by [12] for other redshifts. Predictions of gravity-dark energy coupling model: {\sl full line} --  the fitted acceleration solution (s3); {\sl dashed line} -- the fitted constant expansion solution (s2); {\sl dotted line} --  phase transition predicted by Eq.\,(\ref{eq:f1}) from the reconstructed evolution history of dynamical dark energy $w(z)$ shown in Fig.\,2.
}
\end{center}
\end{figure} \label{fig:h}

\begin{figure}
\begin{center}
\psfig{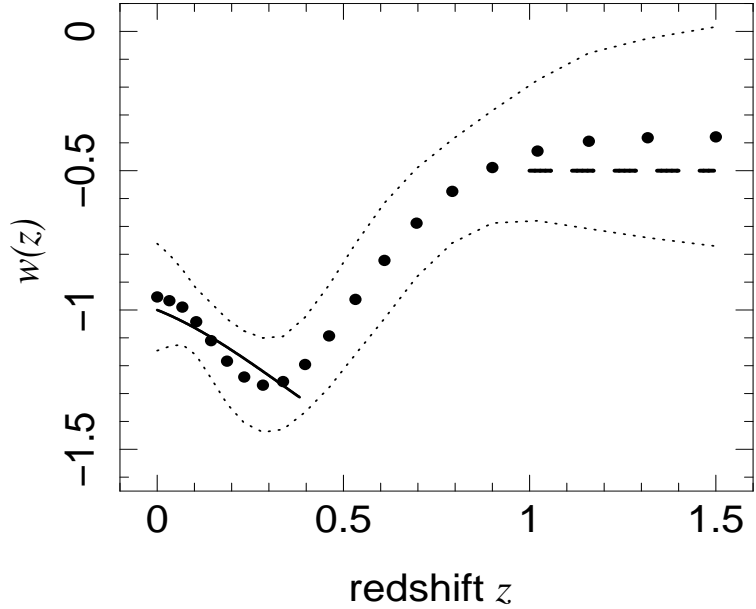}
\caption{Equation of state parameter $w(z)$ of dark energy versus redshift. Dots and dotted lines show reconstructed $w(z)$ and 68\% C.L. error band from \cite{zhao12} (courtesy of G.B. Zhao). Predictions of gravity-dark energy coupling model: {\sl full line} --  the fitted acceleration solution (s3); {\sl dashed line} -- the constant expansion solution (s2).
}
\end{center}
\end{figure} \label{fig:w}

  Now we test the model with cosmological data. The observationally measured data of Hubble parameter is invaluable in probing the expansion history of universe$^{\cite{ztj10}}$. The observed data of $H(z)$ between redshifts $0.07\le Z\le 2.3$ from 28 independent measurements$^{[5-11]}$ compiled by Farooq and Ratra$^{\cite{far13}}$ and $H_0=67.3\pm 1.2$\,km/s/Mpc for $z=0$ from {\sl Planck} 2003 results$^{\cite{ade13}}$ are shown in Fig.\,1.  Another useful probe is the equation of state parameter $w(z)$ of dark energy.  Zhao et al.$^{\cite{zhao12}}$ reconstructed the evolution history of $w(z)$ from combining the latest supernova (SNLS3), cosmic microwave background, redshift space distortion, and the baryonic acoustic oscillation measurements, their results are shown in Fig.\,2. 

\begin{figure}[t]
\begin{center}
\psfig{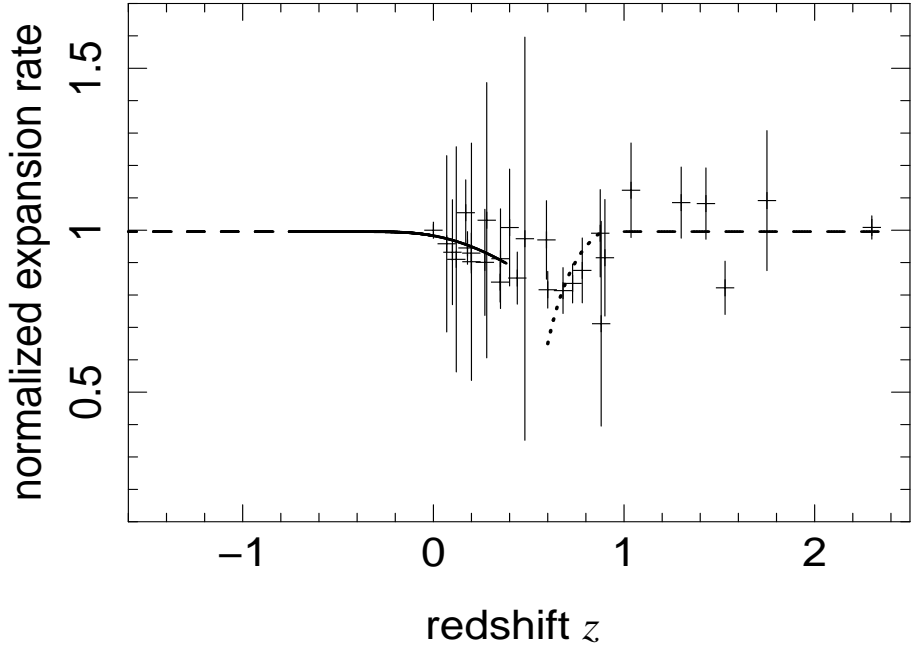}
\caption{Normalized expansion rate $\dot{a}(z)/\dot{a}(0)=H(z)/[(1+z)H_0]$ versus redshift. Predictions of gravity-dark energy coupling model: {\sl full line} --  the fitted acceleration solution (s3); {\sl dashed line} -- the fitted constant expansion solution (s2); {\sl dotted line} --  phase transition predicted by Eq.\,(\ref{eq:f1}) from the reconstructed evolution history of dynamical dark energy $w(z)$ shown in Fig.\,2.
}
\end{center}
\end{figure} \label{fig:ad}

Supernova observations$^{\cite{rie98,per99}}$ show that the expansion of our universe has been speeding up, therefore under our model the present universe is in an acceleration phase describing by the solution (s3). Fig.\,3 shows observed expansion rates deduced from Fig.\,1 with $\dot{a}=H a$. We find that from $z\simeq 0.5$ to the present ($z=0$) the expansion rate has been really increasing; however, earlier universe at redshifts $z\ge 1$ seems not in a deceleration phase expected by conventional standard models: the measured rates are more consistent with being in constant expansion. For a uniformly expanding universe under the solution (s2), we have $\rho_{_d}=0$ and a constant equation of state parameter of dark energy $w(z)=-0.5$ (shown by the left dashed line in Fig.\,2, which is also compatible with observations for dynamical dark energy. Therefore, cosmological observations for $H(z)$ and $w(z)$ indicate that the current acceleration phase of our universe was started from $z\simeq 0.5$, and the early universe at redshifts $z>1$ was in a constant expansion phase, between the two phases the universe was going through a phase transition.

The solutions (s1)-(s3) can not be used to describe the universe during phase transition, because the physics of phase transition is not included in the evolution equation\,(\ref{eq:f3}). However, The universe scale $a$, rate $\dot{a}$ and differential energy density $\rho_{_d}$ should satisfy the energy conservation equation (1) during the whole expansion.  From the observed variation history of $w(z)$ in the region of $z$ between 0.6 and 0.9 shown in Fig.\,2, using $\rho_{_d}=(1+2w)\rho_0$ we can derive the evolution history of differential density during phase transition. With the transition history of $\rho_{_d}$,  expected Hubble parameters $H(z)$ during the phase transition can be evaluated by the energy conservation equation (\ref{eq:f1}). 

\begin{figure}
\begin{center}
\psfig{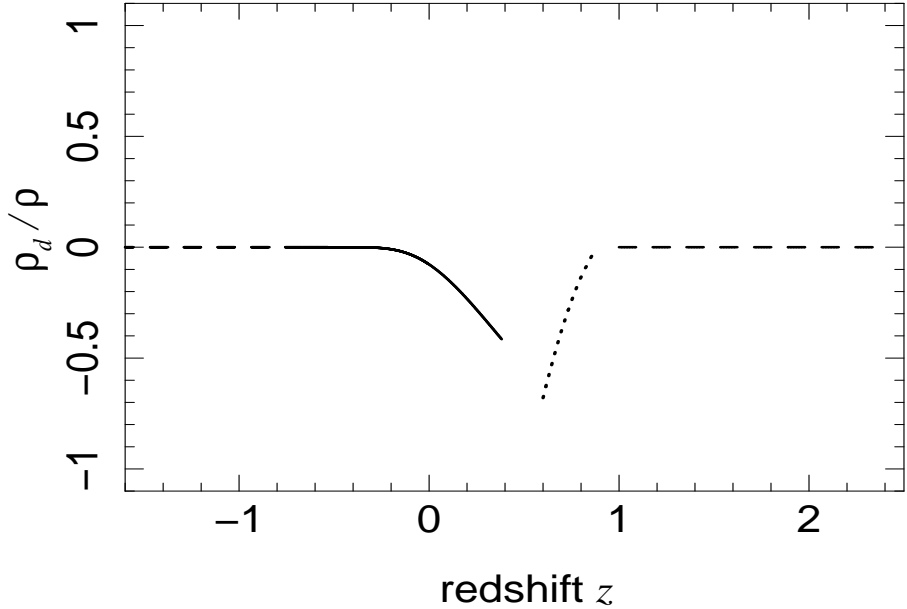}
\caption{Evolution of fractional differential density $\rho_{_d}(z)/\rho(z)$:  after a gravity-dark energy equilibrium period at $z\ge 1$ ({\sl dashed line}), the universe entered a phase transition epoch while  gravitational energy was being converting into dark energy ({\sl dotted line}). The present dark energy dominated acceleration period started at $z\simeq 0.5$ and will continue to $z=-0.7$ ({\sl full line}), then enter into a next  gravity-dark energy equilibrium period ({\sl dashed line} at $z<-0.7$).
}
\end{center}
\end{figure} \label{fig:rd}

 Taking the rest energy density $\rho_{_0}$ of the present universe and constants $\epsilon$, $\mu$ and $c_2$ in (s2) and (s3) as four unknown parameters to be determined, we fit the acceleration expansion solution (s3) at $z\le 0.4$ and the constant expansion solution (s2) at $z\ge 1$ to observed data of both $H(z)$ and $w(z)$, and fit Hubble parameter values predicted by $w(z)$ at $z\in (0.6,0.9)$ to observed $H(z)$ simultaneously. The relationships $a(z)=(1+z)^{-1}$, $H(z)=\dot{a}(z)/a(z)$, $w(z)/w(0)=\rho_{_\lambda}(z)/\rho_{_\lambda}(0)$, $\rho_{_\lambda}=(\rho-\rho_{_d})/2$ and Eq.\,(\ref{eq:f1}) are used in fitting. The resulting evolution histories of $H(z)$ and $w(z)$ are shown in Figs.\,1 and 2. The model predictions fit observed data quite well with reduced $\chi^2=0.83$. The evolution history of differential density $\rho_{_d}$ is shown in Fig.\,4, where $\rho_{_d}$ are evaluated by Eq.\,(\ref{eq:f1}) from the fitted solutions (s2) and (s3) for constant and acceleration epochs and by $\rho_{_d}(z)=[1+2w(z)]\rho_{_0}$ with reconstructed $w(z)$ shown in Fig.\,2 and fitted $\rho_{_0}$.

Based on observations for dynamical Hubble parameter and dark energy, the gravity-dark energy coupling model presents a unified and self-consistent picture for the evolutionary history of the universe: our universe has been going through three evolutionary epochs: constant expansion at $z>1$, phase transition at $z\in (0.5, 1)$, and acceleration starting from $z\simeq 0.5$. The expected deceleration epoch after inflation has not been detected yet, we hope that future deep observations of the earlier ages will find the starting point of the observed constant expansion to know how long the gravity-dark energy balanced period had lasted, and then determine the integral constant $c_1$ of the solution (s1). With the solution (s1) and the conservation of mechanical energy, we will be able to  describe the matter dominated deceleration epoch from its beginning. 

The physical laws and gravity-dark energy coupling which governs the evolution of the observed universe could also be used to speculate the very early universe. With energy conservation, the energy density $\rho(z)=\rho_{_0} a^{-3}=\rho_{_0} (1+z)^{3}$ is increasing as time is going back. The vacuum energy density $\rho_{_V}$ predicted by the uncertainty principle sets an upper limit for $\rho(z)$.  From the so called "the cosmic constant problem"$^{\cite{wei89}}$ we know  that $\rho_{_V}/\rho_{_0}\simeq 10^{120}$, consequently $z_s\simeq 10^{40}$ should be an upper limit for redshift, otherwise the energy density of our universe would be larger than the zero-point energy. Therefore, our universe could be generated with an initial energy density $\rho_{_V}$ at $z_s\simeq 10^{40}$ from a phase transition in a limited region of $R_s\simeq 10^{-40} R_0$  within a static vacuum with balanced attraction-repulsion fields$^{\cite{li11}}$. The primordial phase transition broke the equilibrium of the static primordial vacuum and resulted in inflation.

From the fitted parameters $\rho_{_0}$, $\epsilon$, and $\mu$, we can derive the radius $R_0$ and mechanical energy density $\rho_{_{mech,0}}$ of the present universe.  Their errors can be estimated by Monte-Carlo simulations with simulated data sampled from observed $H(z)$ or $w(z)$ data.  The fitted solution (s3) for present universe can also help us to predict future evolution of our universe. Calculating $\dot{a}(z)$ and $\rho_{_d}(z)$ with the fitted solution (s3) and Eq.\,(1) at $z<0$ we find that $\dot{a}$ will increase up to be equal to the critical rate $\dot{a}_c$  at $z=-0.7$  while $\rho_{_d}$ to be zero (see the region of $z<0$ in Figs.\,3 and 4), i.e. in the near future our universe will cease to accelerate and return to the gravity-dark energy balanced state again.

A gravity-dark energy coupled universe can remain in an equilibrium state with constant expansion rate. Inspecting Figs.\,3 and 4 we can see that, it is really possible that constant expansion is ordinary for our universe, phase transition and acceleration (or deceleration) are just temporary deviation from and succeeding relaxation to equilibrium state. Phase transitions with transformation between attractive  and repulsive fields powered the generation and evolution of our universe. The primordial phase transition occurred in static vacuum partially converted rest energy into mechanical energy under conservation of total energy. A remarkable finding here is a phase transition just occurring at $z\in (0.5, 1)$ while the gravitational energy was converted into dark energy under conservations of rest energy and mechanical energy separately.  The acceleration of present universe can be seen as an ongoing inflation, which is similar to the primordial inflation, both are resulted from phase transition, powered by dark energy, and last similar period length in logarithmic scales. Therefore, the dynamic era from $z\simeq 1$ to present of our universe is active and interesting not only for astrophysics, but also for cosmology, and for fundamental physics as well.

\noindent{\bf Acknowledgments}\\
\noindent This work is supported by the National Natural Science Foundation of China (Grant No. 11033003).\\

\end{document}